\documentclass[prb,aps,amsmath,amssymb,superscriptaddress,citeautoscript,preprint]{revtex4-1}
\usepackage[utf8]{inputenc}
\usepackage[english]{babel}
\usepackage{graphicx}  
\usepackage{dcolumn}   
\usepackage{bm}        
\usepackage{verbatim}   
\usepackage{color}		
\usepackage{lipsum}

\makeatletter
\newcommand*{\citenums}[2][]{%
	\begingroup
	\let\NAT@mbox=\mbox
	\let\@cite\NAT@citenum
	\let\NAT@super@kern\relax
	\renewcommand\NAT@open{}%
	\renewcommand\NAT@close{}%
	\cite[#1]{#2}%
	\endgroup
}
\makeatother

\begin{document}
	
	\title{Formation dynamics of exciton polariton vortices created by non-resonant annular pumping}
	\author{Bernd Berger}
	\author{Daniel Schmidt}
	\affiliation{
		Experimentelle Physik 2,
		Technische Universit\"at Dortmund,
		D-44221 Dortmund, Germany
	}
	
	\author{Xuekai Ma}
	\affiliation{
		Department of Physics and Center for Optoelectronics and Photonics Paderborn (CeOPP), Universit\"at Paderborn, Warburger Strasse 100, 33098 Paderborn, Germany
	}
	
	\author{Stefan Schumacher}
	\affiliation{
		Department of Physics and Center for Optoelectronics and Photonics Paderborn (CeOPP), Universit\"at Paderborn, Warburger Strasse 100, 33098 Paderborn, Germany
	}
	\affiliation{College of Optical Sciences, University of Arizona, Tucson, AZ 85721, USA}
	
	\author{Christian Schneider}
	\affiliation{Technische Physik, Physikalisches Institut and W\"{u}rzburg-Dresden Cluster of Excellence ct.qmat, Universit\"{a}t W\"{u}rzburg, Am Hubland, 97074, W\"{u}rzburg, Germany}
	
	\author{Sven H\"{o}fling}
	\affiliation{Technische Physik, Physikalisches Institut and W\"{u}rzburg-Dresden Cluster of Excellence ct.qmat, Universit\"{a}t W\"{u}rzburg, Am Hubland, 97074, W\"{u}rzburg, Germany}
	\affiliation{SUPA, School of Physics and Astronomy, University of St. Andrews, St. Andrews KY16 9SS, United Kingdom}
	
	\author{Marc A\ss mann}
	\affiliation{
		Experimentelle Physik 2,
		Technische Universit\"at Dortmund,
		D-44221 Dortmund, Germany
	}
	
	\date{\today}
	\begin{abstract}
		We study the spontaneous formation of exciton polariton vortices in an all-optical non-resonantly excited annular trap, which is formed by ring-shaped subpicosecond laser pulses. Since the light emitted by these vortices carries orbital angular momentum (OAM) corresponding to their topological charge, we apply a dedicated OAM spectroscopy technique to detect the OAM of the measurement signal with picosecond time resolution. This allows us to identify the formation of OAM modes and investigate the dynamics of the vortex formation process. We also study the power dependence of this process and how the ring diameter influences the formation of OAM modes.
	\end{abstract}
	
	\maketitle
	
	\section{Introduction}
	
	Exciton polaritons provide an ideal platform for all-optical information processing due to their direct optical accessibility, which allows one to control and use non-linear interactions between exciton polaritons. A very prominent non-linear effect is stimulated polariton-polariton scattering enabling Bose-Einstein condensation (BEC) of exciton polaritons \cite{kasprzak2006}. One of the first demonstrations included the buildup of a spatially coherent polariton condensate using a non-resonant excitation laser.
	Subsequently the formation of condensates inside natural traps\cite{sanvitto2009}, fabricated traps\cite{Balili_2007,Bajoni_2008,Winkler_2015}, and optically imprinted trapping potentials has been studied\cite{wertz_2010,askitopoulos_2013,askitopoulos_2016}. Also tailored flow control\cite{Sanvitto2011, assmann_2012,schmutzler_2015,winkler_2017} and pattern formation\cite{Tosi_2012,Dreismann_2014,Whittaker_2017} of polaritons in 2D microcavities were shown. In this work we focus on observing the formation of exciton polariton vortices, which feature a $2\pi\cdot m$ radial phase change of the polariton wavefunction with the topological charge $m$ and a phase singularity in the centre of the vortex corresponding to a minimum in the polariton density function, i.e. the core of the vortex. Various ways of exciting exciton-polariton vortices have been investigeted experimentally\cite{lagoudakis2008,nardin_2010,Sanvitto_2010,Lagoudakis_2011,Dall_2014,Boulier2015,Boulier_2016,Kalinin_2017,Dominici2018,kwon_2019} and theoretically\cite{Liew_2007,Boulier2015,Boulier_2016,Ma_2016,Ma_2017,Ma_2018}. Also vortex lattices\cite{Liew_2008,Keeling_2008,Tosi2012a,Hivet_2014,Ohadi_2016} and vortex pairs\cite{Fraser_2009,Roumpos2010,Nardin2011,Manni_2013} have been studied.
	
	Here, we experimentally and theoretically study the spontaneous formation dynamics of vortices in an optically created annular trap. Since the topological charge of an exciton-polariton vortex in a microcavity directly maps to the OAM of the emitted light, we apply a dedicated OAM spectroscopy technique, which allows us to resolve individual OAM states\cite{Berger:18}.

	\section{Experimental Methods}
	
	The sample we employ to investigate exciton-polaritons is a MBE-grown planar microcavity based on GaAs. It has a quality factor of about 20.000 and a Rabi splitting of 9.5\,meV and consists of two distributed Bragg reflectors (DBR) made of 32 and 36 alternating layers of $\text{Al}_{0.2}\text{Ga}_{0.8}\text{As}$ and AlAs enclosing a $\lambda/2$ cavity. Four GaAs quantum wells are placed in the central antinode of the cavity light field. The sample is mounted on the cold finger of a helium-flow cryostat, which cools it down to a temperature of 17\,K. The exciton-cavity detuning is -4\,meV for all measurements shown in this work.
	
\begin{figure}
		\includegraphics{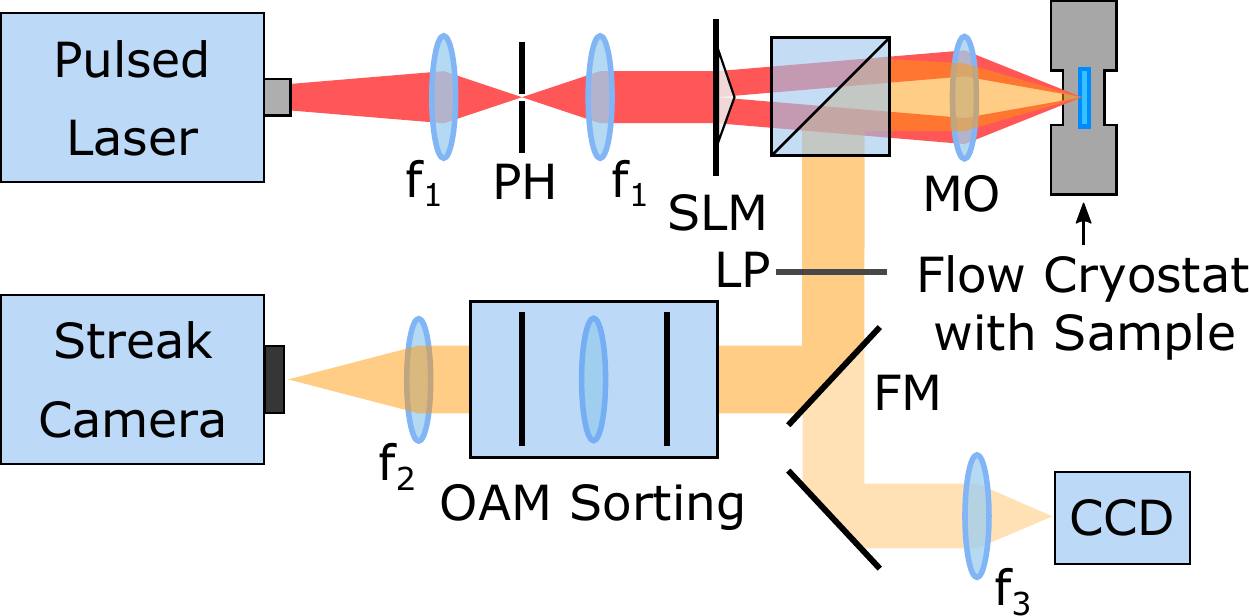}
		\caption{Simplified sketch of the Setup. The SLM displays an axicon phase with variable slope and shapes the beam to rings. The MO is 20x with 0.40 NA, $f_1$=100\,mm and $f_2$=190\,mm,  $f_3$=750\,mm, PH a is 50\,$\mu$m pinhole and LP is a longpass filter at 750\,nm. The OAM sorting part consists of the same parts as in Ref. [\citenums{Berger:18}]: Transform and Correction phase pattern on another SLM and a f=100\,mm concave mirror. For simplicity all phase patterns imprinted on the beam are depicted as transmissive elements, while our SLM works in reflection geometry. Polarization optics are omitted. Also not shown is the reduction of the signal beam diameter by a factor of 3 before entering the OAM sorting.}
		\label{fig:SetupSketch}
	\end{figure}

	The experimental setup implements OAM spectroscopy\cite{Berger:18} in the detection part of the setup. A sketch is shown in Fig. \ref{fig:SetupSketch}. For non-resonant optical excitation a pulsed titanium-sapphire laser (with repetition rate of 75.39\,MHz) emitting pulses with a duration of approximately 120\,fs at a central wavelength of 735.5\,nm (1686\,meV) is used. The pulsed laser beam is shaped using a spatial light modulator (SLM) to generate an annular optical potential in the focal plane, where the sample is located. The shaped beam is focused onto the sample using a microscope objective (numerical aperture 0.4). This microscope objective also collects the polariton emission, which is filtered with a long-pass filter to remove the reflection of the excitation laser beam. The real space image then may be acquired by directing the light to a liquid nitrogen-cooled CCD camera placed behind a monochromator (operated in zeroth order). Additionally, guiding the signal beam trough the OAM sorting process to a streak camera allows for time resolved mapping of the OAM. In this case, the horizontal axis of the measured image maps the OAM of the signal beam while the vertical axis maps the time relative to the arrival of the pump beam with few picosecond resolution.
	
	The OAM sorter in our setup works as follows: The signal beam carrying OAM is imaged onto an arrangement of two custom phase elements, which are displayed on two halves of a SLM. First, the beam is imaged onto the transformation phase pattern on the first half of the SLM. It unwraps the helical phase gradients of OAM states into linear phase gradients. However, with only a single phase pattern the transformation does not result in perfectly linear phase gradients. To achieve this, the light beam is collected with a concave mirror, which reflects the light back to the other half of the SLM and performs an optical Fourier transformation of the light beam, like a lens would do. For correction of the remaining deviations of the wave front from linear phase gradients a second phase pattern is placed in the Fourier plane. Finally, the light beam transformed by the OAM sorter is imaged onto a detection device such as a CCD camera or the entrance slit of a streak camera using another lens. OAM states with different phase gradients then are imaged onto corresponding spots at different lateral positions in the plane of detection, so that every OAM state is mapped to one detector position\cite{Berger:18}. The streak camera hereby allows one to perform time resolved OAM measurements and also measurements of correlations between different OAM modes\cite{ASSmann:10}.
	
	\section{Vortex dynamics}
	
\begin{figure}
		\includegraphics{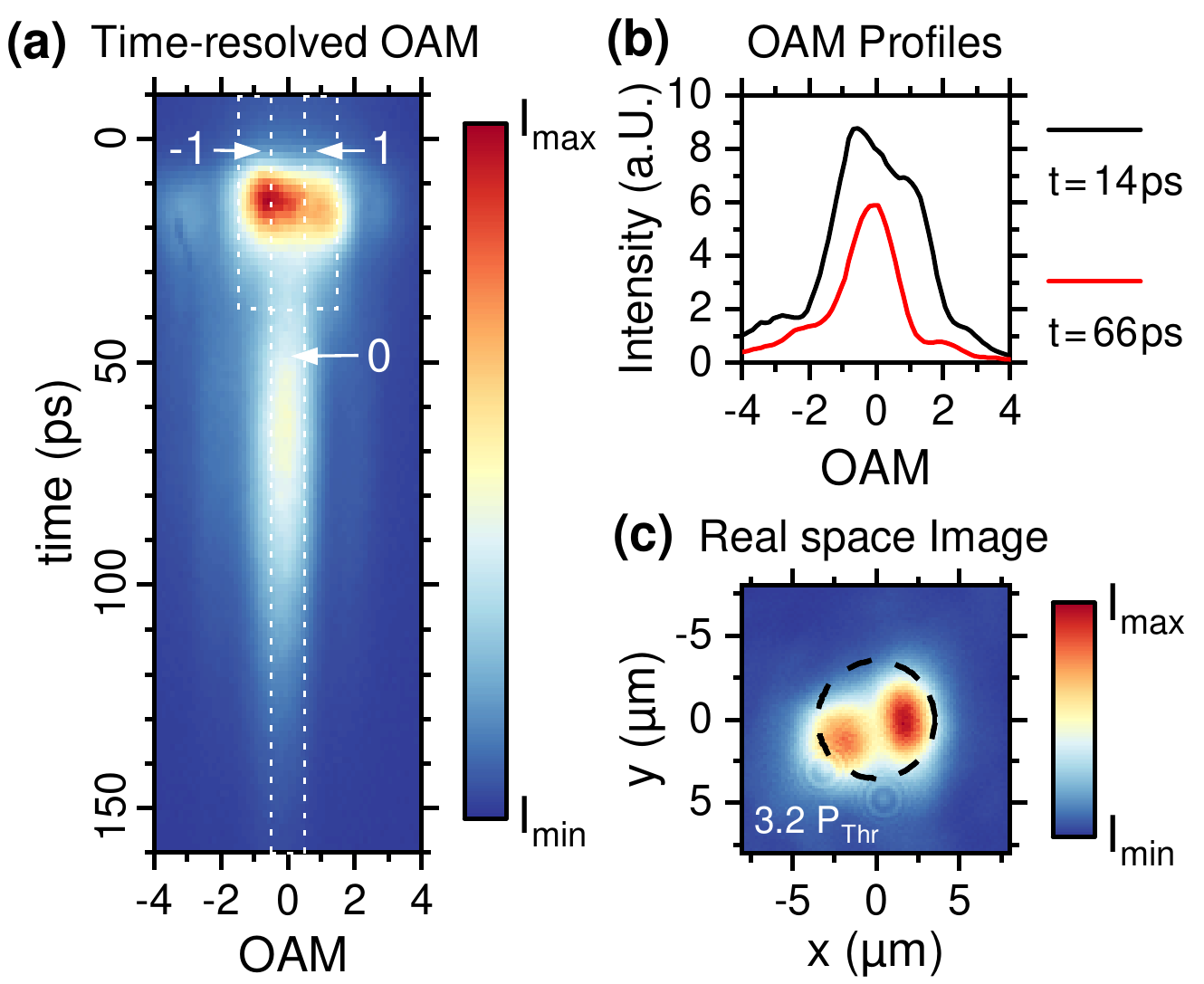}
		\caption{(a) The detected time-resolved OAM signal for a pump ring with 7\,$\mu$m diameter and a pump power of 3.2\,$\text{P}_\text{Thr}$. A characteristic OAM pattern forms which can be segmented into three areas, containing different OAM modes. These areas are used for correlation measurements. (b) Profiles along the OAM axis, obtained by integrating a time interval of $\pm10.5$\,ps around the mode peaks. (c) Real space image of the PL signal above threshold for a 7\,$\mu$m diameter pump ring, located at the position indicated by the black dotted circle. }
		\label{fig:VortexResult}
	\end{figure}

	First we demonstrate that vortices form spontaneously inside the annular potential. In Fig. \ref{fig:VortexResult} typical results of real space and OAM resolved measurements are shown. In real space a dipole-like pattern forms inside the annular pumping potential, which suggests that the polaritons are confined to the annular potential and form localized modes. However, this does not necessarily imply the existence of vortices. Thus, we apply time-resolved OAM spectroscopy, which allows us to confirm the formation of OAM states. We find the polariton condensate to form -1 and +1 OAM states initially and an additional state with zero OAM at a later time. This provides evidence that inside the annular trapping potential indeed vortices are spontaneously created. We find the sample disorder to be of minor importance for this formation process, as the OAM states and time scales of the formed condensates do not vary significantly with different sample positions. Still, the sample disorder may influence the intensity ratio of the -1 and +1 OAM states, just like an asymmetry of the annular pumping potential would do.
	
	The appearance of states with both topological charges -1 and +1 raises the question if vortices of opposite topological charges coexist with each other or if only either of them can exist at the same time. We cannot conclude this from the image shown in Fig. \ref{fig:VortexResult}b since it is integrated over several million single pulse excitations and gives no insight into the individual single pulse events.
	To study the interdependence of the arising states with -1, +1, and zero OAM, we perform correlation measurements\cite{ASSmann:10} and investigate the cross-correlation between the areas indicated by the white boxes -1, +1, and 0 as shown in Fig. \ref{fig:VortexResult}b. The correlations are calculated as follows:
	
	\begin{align}
	C_{n,m}&=\frac{<I_{n}\cdot I_m>}{<I_{n}><I_m>}\\
	A_{n}&=\frac{<I_{n}\cdot (I_{n}-1)>}{<I_{n}>\cdot <I_{n}>}
	\end{align}
	
	Here $C_{n,m}$ is the cross-correlation, $A_{n}$ is the autocorrelation and $n,m \in \{-1,0,1\}$ are indices for the OAM. $I_{n}$ and $I_{m}$ are the respective photon counts recorded for the individual OAM modes.	
	When the obtained cross-correlation value $C$ is $0\leq C < 1$ for a pair of two selected modes, this is evidence that each of these modes may suppress the other one. In this case they will not coexist. However when $C\geq1$ the modes may coexist and for $C>1$ even show the tendency to appear in a correlated manner while coexisting. Also the autocorrelation of all detected photon events in the area of each individual OAM mode is measured to obtain a measure for the photon number noise of the signal. 
	
	As a result, the cross-correlation of the -1 and +1 OAM modes yields a value of $C_{-1,1}=1.08 \pm 0.09$, while the autocorrelation values representing the photon number noise of both individual modes are $A_{-1}=1.15 \pm 0.09$ and $A_{1}=1.2 \pm 0.1$. Since the cross-correlation value is above one and does not significantly differ from the autocorrelation values, we conclude the OAM modes to be essentially statistically independent of each other with a common noise source. So both modes seem to appear randomly and do not influence the occurence of each other, i.e. they coexist with each other.
	
	Further analysis shows that also no cross-correlation between each -1 and +1 OAM mode and the zero OAM mode is found. The values of $C_{-1,0}=1.05 \pm 0.06$ and $C_{1,0}=1.23 \pm 0.07$ are above one and roughly match with the individual noise of both OAM -1 and +1 modes. This implies that the long signal with zero OAM does not directly result from a decay of the OAM modes, but rather is created by a different and independent process.
	
	One key influence factor on the dynamics of polariton condensates is the excitation power used. All arising modes are influenced significantly by the non-resonant pumping power, as shown in Fig. \ref{fig:2D_Plots}. Both experiment and theory (details on the theoretical modelling are given below) show remarkably good qualitative agreement with each other. For low pumping power only slightly above the condensation threshold, only the zero OAM condensate arises at later times while modes with finite OAM are completely absent. With increasing pumping power, the $-1$ and $+1$ OAM condensates also appear simultaneously significantly before the zero OAM mode is formed and the onset of all condensate components shifts to earlier times. The zero OAM condensate shows a stronger shift in time compared to the finite OAM modes when the pumping power is increased. The timescale of the decay of the zero OAM mode, however, is not modified as the decay times do not change significantly for higher excitation powers. This holds true both for experiment and theory.
	
\begin{figure}
		\includegraphics{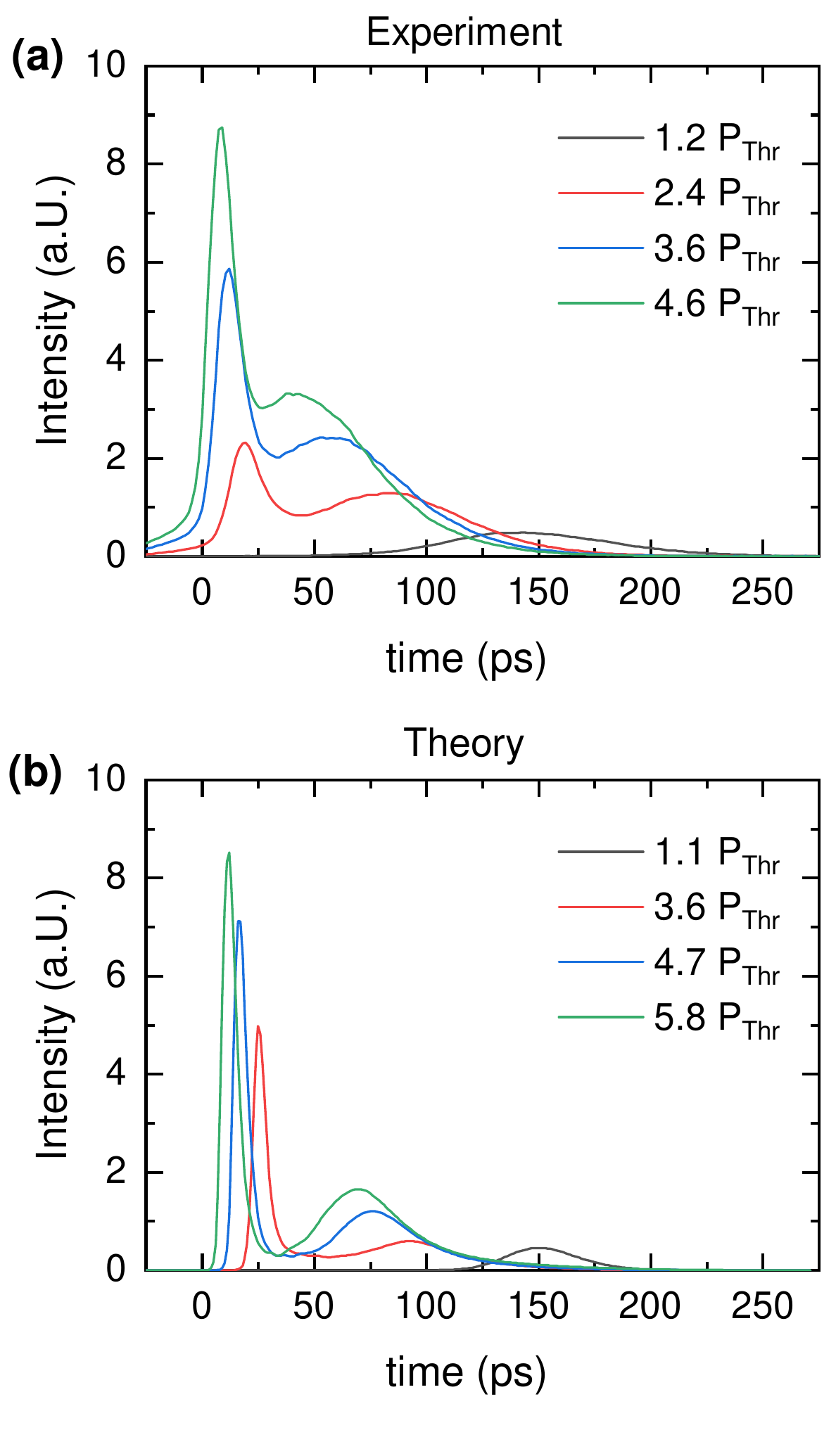}
		\caption{OAM-integrated time profiles in experiment (a) and theory (b) for different excitation powers}
		\label{fig:Profiles}
	\end{figure}

	The OAM-integrated time profiles shown in Fig. \ref{fig:Profiles} allow us to quantify the decay times of the -1 and +1 OAM modes and the zero OAM mode. In experiment, the decay time of the second peak initially first rises with increasing power and then stabilizes at 24\,ps for excitation powers well above twice the condensation threshold.
	The decay time of the first peak corresponding to the -1 and +1 OAM modes amounts to about 12.5\,ps and varies less than 1\,ps in the power range above threshold covered in Fig. \ref{fig:Profiles}. In order to understand the underlying physics of the different time scales of the zero and non-zero OAM modes, we compare the experimental results to a theoretical model, which is introduced in the next section.
	
	\begin{figure}
		\includegraphics{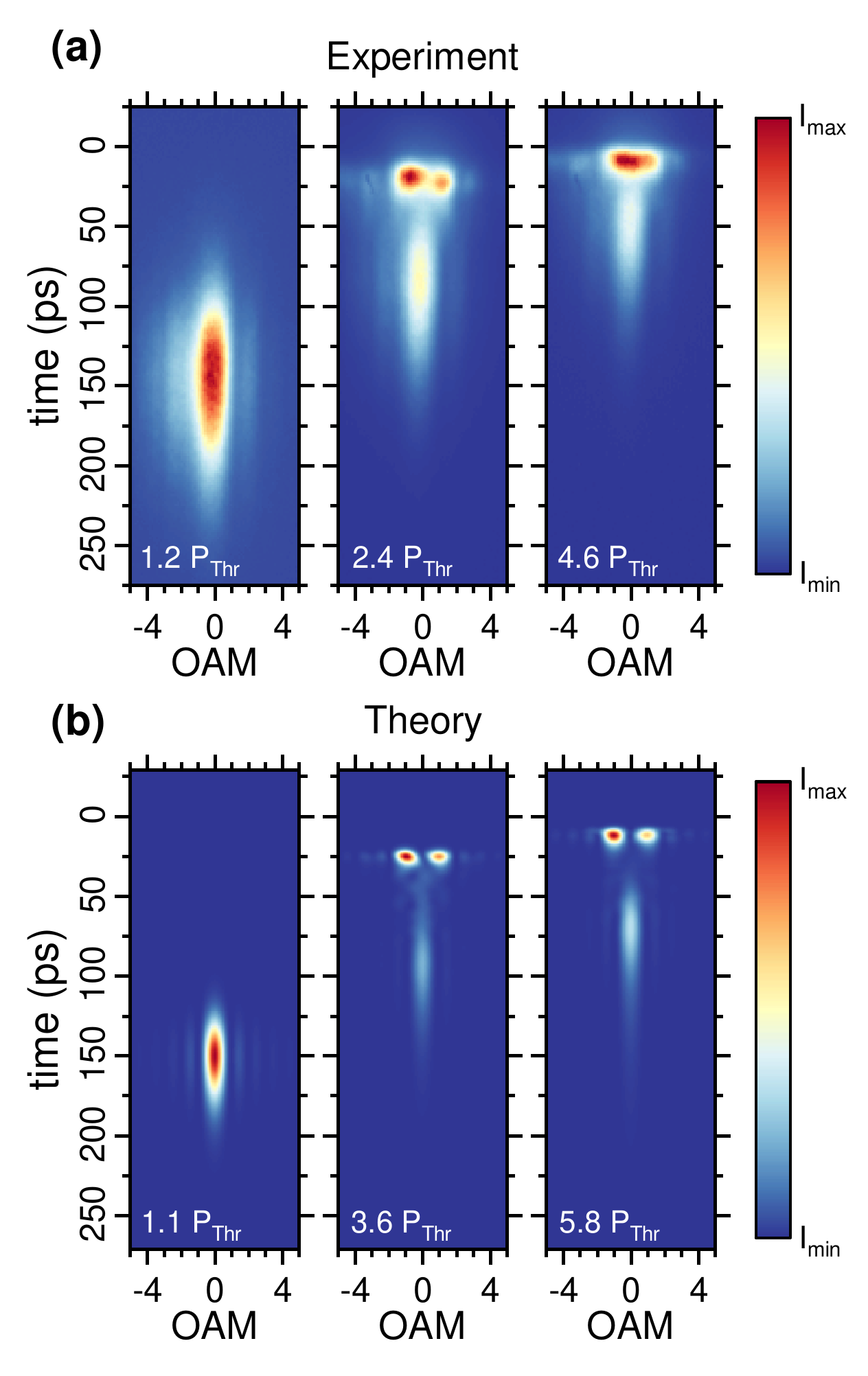}
		\caption{Time-resolved OAM measurements for different excitation powers in experiment (a) and theoretical simulation (b) of the whole process. The colormaps are scaled to the individual maximum and minimum values of the images.}
		\label{fig:2D_Plots}
	\end{figure}
	
	\section{Theory}
	
	The dynamics of polariton condensates in the vicinity of the bottom of the lower polariton branch can be described by a mean-field open-dissipative Gross-Pitaevskii (GP) equation~\cite{wouters2007excitations}:
	\begin{equation}\label{e1}
	\begin{aligned}
	i\hbar\frac{\partial\Psi(\mathbf{r},t)}{\partial t}=&\left[-\frac{\hbar^2}{2m_\text{eff}}\nabla_\bot^2-i\hbar\frac{\gamma_\text{c}}{2}+g_\text{c}|\Psi(\mathbf{r},t)|^2 \right.\\
	&+\left.\left(g_\text{r}+i\hbar\frac{R}{2}\right)n_\text{A}(\mathbf{r},t)+{g_\text{r}}n_\text{I}(\mathbf{r},t)\right]\Psi(\mathbf{r},t),
	\end{aligned}
	\end{equation}
	where $\Psi(\mathbf{r},t)$ is the wavefunction of the polariton condensate. The effective mass of lower polaritons is represented by $m_\text{eff}=10^{-4}m_\text{e}$ ($m_\text{e}$ is the free electron mass). $\gamma_\text{c}=0.09$~ps$^{-1}$ represents the loss rate of condensates, resulting from the finite lifetime of polaritons. The repulsive polariton-polariton interaction strength is given by $g_\text{c}=2$~$\mu$eV~$\mu\text{m}^{2}$. To describe the relaxation and scattering dynamics under non-resonant excitation, we consider two reservoirs, an active reservoir and an inactive reservoir, for the whole condensation procedure~\cite{lagoudakis2011probing,schmutzler_2015}. The condensates are replenished directly from the active reservoir, $n_\text{A}$, with the condensation rate $R=0.02$~ps$^{-1}$~$\mu\text{m}^{2}$, in a stimulated manner. The interaction between condensates and the active reservoir is represented by $g_\text{r}=4$~$\mu$eV~$\mu\text{m}^{2}$. The density of the active reservoir satisfies
	\begin{equation}\label{e2}
	\frac{\partial n_\text{A}(\mathbf{r},t)}{\partial t}=\tau n_\text{I}(\mathbf{r},t)-\gamma_\text{A} n_\text{A}(\mathbf{r},t)-R|\Psi(\mathbf{r},t)|^2n_\text{A}(\mathbf{r},t).
	\end{equation}
	Here, $\gamma_\text{A}=0.005$~ps$^{-1}$ is the loss rate of the active reservoir. The active reservoir is replenished by the inactive reservoir, $n_\text{I}$, with rate $\tau=0.016$~ps$^{-1}$. The inactive reservoir contains hot excitons excited directly by the non-resonant pump. The density of the inactive reservoir satisfies
	\begin{equation}\label{e3}
	\frac{\partial n_\text{I}(\mathbf{r},t)}{\partial t}=-\tau n_\text{I}(\mathbf{r},t)-\gamma_\text{I} n_\text{I}(\mathbf{r},t)+P(\mathbf{r},t).
	\end{equation}
	Here, $\gamma_\text{I}=0.005$~ps$^{-1}$ is the loss rate of the inactive reservoir. The energy of the non-resonant pump, $P(\mathbf{r},t)$, for which we assume a ringlike intensity profile, is far above the excitonic resonance. In the simulations, the excitation pump has a similar shape and duration to that used in experiments. It has the form 
		\begin{equation}\label{e3}
		P(\mathbf{r},t)={P_0}e^{-\left(\frac{\mathbf{r}}{w_1}\right)^4}\left[1-{c}e^{-\left(\frac{\mathbf{r}}{w_2}\right)^4}\right]e^{-\left(\frac{t}{w_t}\right)^2}.
		\end{equation}
		Here, $P_0$ is the intensity of the pump, $w_1=5$ $\mu$m and $w_2=2.5$ $\mu$m determine the radius of the pump, $c=0.7$ gives rise to the weaker intensity in the center of the ring pump, and $w_t=0.15$ ps represents the duration of the pulse.
	
\begin{figure}
		\includegraphics{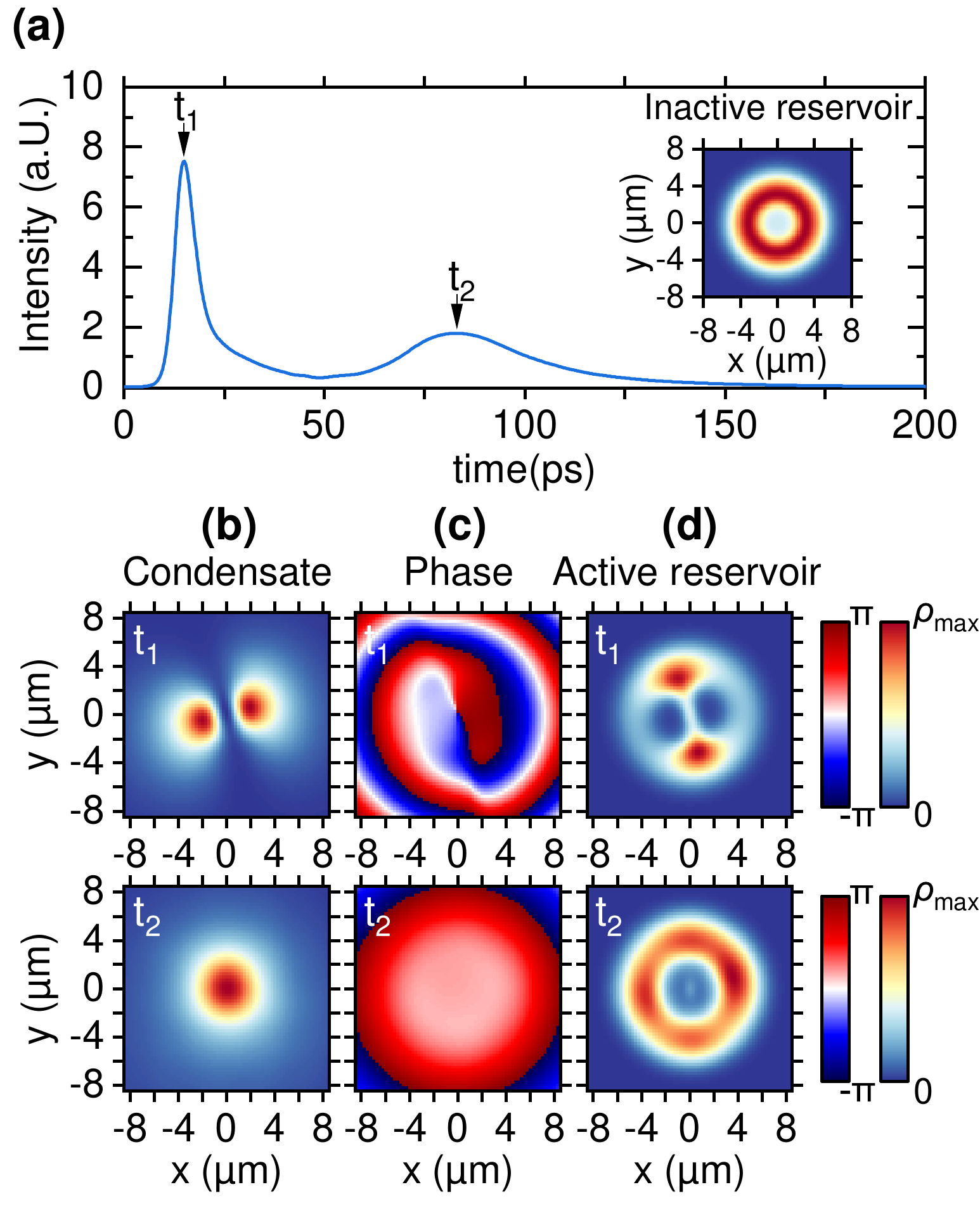}
		\caption{The panel (a) shows a simulated OAM-integrated time profile for a pumping power of 4.7\,P$_{\text{Thr}}$. The corresponding real space image of the condensate, phase map of the condensate and real space image of the active reservoir at the peak times $t_1=15\,$ps and $t_2=83\,$ps are shown in (b), (c) and (d). The inset in (a) shows the persistent shape of the inactive reservoir, which decays and fills the active reservoir. $\rho_{\text{max}}$ at t$_1$ (t$_2$) equals the particle densities 52\,$\mu m^{-2}$ (6.9\,$\mu m^{-2}$) for the condensate and 77\,$\mu m^{-2}$ (17\,$\mu m^{-2}$) for the active reservoir.
		}
		\label{fig:Theory_Realspace}
	\end{figure}

	The numerically sorted intensity distributions ($\sim|A_{\text{OAM}}|^2$) of the OAM presented in Fig. \ref{fig:2D_Plots}(b) are calculated by~\cite{Ma_2020}
	\begin{equation}\label{e4}
	A_{\text{OAM}}(m)=\int\Psi({\bf r})e^{im\phi}d{\bf r},
	\end{equation}
	where, $\phi$ is the polar angle and the origin of the coordinate
	system is located in the center of the ring-shaped excitation
	pulse. $m$ represents the value of the OAM. The data in Figs. \ref{fig:2D_Plots}(b), \ref{fig:Profiles}(b) and \ref{fig:Theory_Realspace} are obtained by averaging over 10 pulsed excitation cycles, while randomly changing the initial white noise in both phase and amplitude for each excitation. The small number of averaged excitation cycles gives rise to the asymmetry of the -1 and +1 OAM states as shown in Fig. \ref{fig:2D_Plots}(b), which is coincidental with the experimental results. Note that if more excitation pulses are averaged, a symmetric distribution of the -1 and +1 OAM states would be expected. For the plots in Fig. \ref{fig:Theory_Realspace}, we select one of the pumping powers shown in Fig. \ref{fig:Profiles}(b), 4.7\,$\text{P}_\text{Thr}$, and extract the real space images of the inactive reservoir, the active reservoir and condensate in a second run of the theory simulation with the same set of parameters as before. Although the initial conditions are not exactly the same due to the noise, still the OAM-integrated time profiles in Fig. \ref{fig:Theory_Realspace}(a) and \ref{fig:Profiles}(b) strongly resemble each other.
	
	The two reservoirs act as repulsive potentials for the polaritons, which can be seen by comparing Eq. \eqref{e1} to the Schr\"odinger equation. The terms $g_\text{r}n_\text{A}$ and $g_\text{r}n_\text{I}$  take the same role as a potential in the latter.
	Therefore the reservoirs carry the same ring shape as the pump before the condensate forms. When the condensate forms, the active reservoir changes its shape due to the condensation process according to the non-linear scattering term $R|\Psi(\mathbf{r},t)|^2n_\text{A}(\mathbf{r},t)$ in Eq. \eqref{e2}. However, the shape of the inactive reservoir is not influenced by condensate formation, since it does not experience feedback from the active reservoir or condensate according to Eq. \eqref{e3}.
	
	The excitation power directly influences the potentials. When the excitation power is low, the resulting potentials are weak. In this case only the fundamental mode with zero OAM forms [see the left panels in Figs. \ref{fig:2D_Plots}(a) and \ref{fig:2D_Plots}(b), and Fig. \ref{fig:Profiles}(b)]. With higher excitation power the reservoir potentials become stronger, which enables the formation of a dipole mode [Fig. \ref{fig:VortexResult}(a)]. As the reservoir density decays after the pulsed excitation, the reservoir potentials become weaker and the dipole mode becomes unstable. Similar to the case with lower excitation power at this time only the fundamental mode with zero OAM forms [Figs. \ref{fig:2D_Plots} and \ref{fig:Profiles}].
	
	For a better illustration, we pick one of the excitation powers corresponding to the OAM-integrated time profiles in Fig. \ref{fig:Profiles}(b) and simulate the theoretical real space images of the reservoirs and condensate. The results for such a typical condensation process are shown  in Fig. \ref{fig:Theory_Realspace}. Fig. \ref{fig:Theory_Realspace}(a) shows the OAM-integrated emission. At the peak times $t_1$ and $t_2$ where the different OAM modes reach the maximum of their occupation, the density and phase of the condensate and the active reservoir density are extracted and shown in Figs. \ref{fig:Theory_Realspace}(b), \ref{fig:Theory_Realspace}(c) and \ref{fig:Theory_Realspace}(d).
		At $t_1$ where the -1 and +1 OAM modes build up, the condensate is forming a dipole-like mode that consists of two in real space localized spots with a $\pi$ phase jump to each other. This can be considered as the result of a coherent superposition of the two counter rotating vortex modes with -1 and +1 OAM, leading to a standing wave.
		At later time $t_2$, when the non-zero OAM modes have decayed and the zero OAM mode arises, the condensate forms a single spot with a flat phase profile in the centre of the excitation ring.
		Regarding the active reservoir density presented in \ref{fig:Theory_Realspace}(d), it becomes apparent that the active reservoir inherits the ring shape of the inactive reservoir and additionally shows local local minima at the condensate locations, since at these spots the polariton population is transferred from the active reservoir into the condensate.
	
	From Eq. \eqref{e1} one can see that the decay rate of the condensate depends on both its lifetime and the gain from the active reservoir, i.e. $-i\hbar\frac{\gamma_\text{c}}{2}+i\hbar\frac{R}{2}n_\text{A}(\mathbf{r},t)$. The density profiles of the condensate and the active reservoir at $t_2$ shown in Fig. \ref{fig:Theory_Realspace} illustrate that for the fundamental mode $n_\text{A}\Psi\simeq0$ due to their negligible overlap, resulting in the decay of the zero OAM signals in Fig. \ref{fig:Profiles} being independent of the condensate density (or pumping power). Therefore, the lifetime of the condensate solely determines the decay of the fundamental mode, i.e. $2/\gamma_\text{c}\simeq22$ ps. The first peak decays much faster than the second one, because the dipole mode becomes unstable very quickly as the potential reduces with shrinking population of the active and inactive reservoir.
	
	For the decay of the first peak in Fig. \ref{fig:Profiles} the theory predicts a value of only 4\,ps, whereas the value in experiment is around 12.5\,ps. So in experiment the early -1 and +1 OAM processes are about three times longer than expected from theory while the decay time of the zero OAM process matches the predictions from theory. We ascribe this deviation to the intensity fluctuation of the ringlike excitation laser beam, which is inevitably induced by the SLM used for beam shaping. Since the entire process is strongly power dependent, this leads to a noticeable broadening in the time domain. This results in a such different values for the first peak decay times in experiment and theory. Also an unavoidable time jitter of the streak camera system additionally broadens the peaks in experiment, however, this is only about 1\,ps and certainly not the main contribution to the broadening in this case.
	
	\section{Influence of ring diameter}
	
\begin{figure}
		\includegraphics{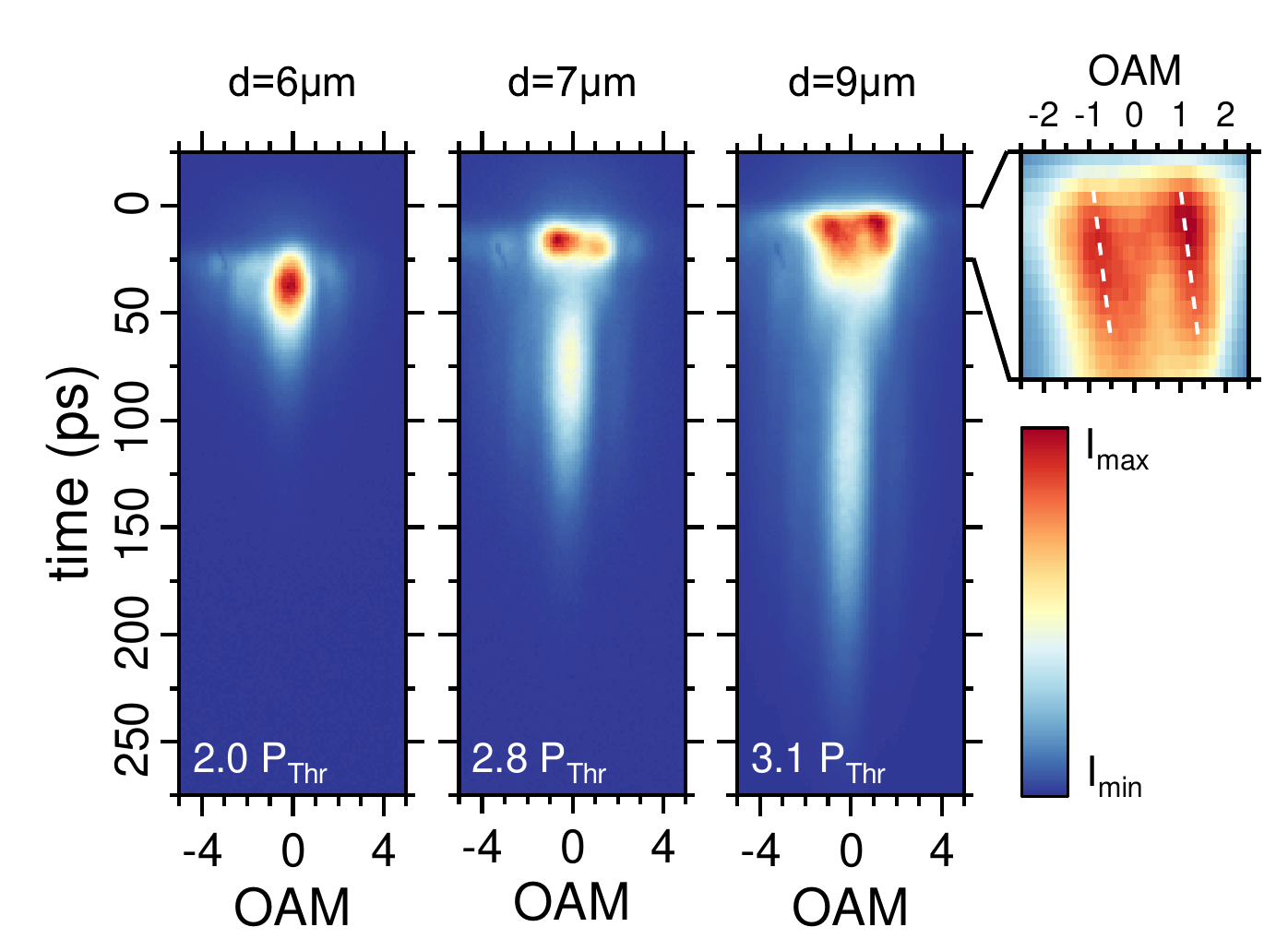}
		\caption{Time-resolved OAM measurements for excitation rings with different diameters but with roughly the same excitation power density. The colormaps are scaled to the individual maximum and minimum values of the images, the pump power relates to the individual thresholds for different ring diameters.
		}
		\label{fig:Different_Rings}
	\end{figure}

	We also study the influence of the ring diameter on the formation of the condensates, as shown in Fig. \ref{fig:Different_Rings}. Here, the ring shaped pump simultaneously creates the polariton reservoir and the trapping potential. We observe significant changes of all modes with different OAM, when tuning the ring diameter while adjusting the power to keep the excitation power density approximately the same.
	When comparing the OAM signals for different ring diameters we find the -1 and +1 OAM modes to form only for ring diameters above 6\,$\mu$m, while the zero OAM mode exists for all presented ring diameters. For a ring diameter of 7\,$\mu$m the vortex modes with non-zero OAM form spontaneously and decay rapidly. When the ring diameter is much larger, for example 9\,$\mu$m, the vortex modes also are formed. However, the signal of both vortex modes decays slower and the spots are moving by approximately one third of a topological charge towards more positive OAM during this decay.
	
	In contrast the zero OAM mode does not move along the OAM axis. It only shifts to a significantly later time with increasing ring diameter and strongly broadens in time. This can be explained by the changing spatial overlap between the zero OAM mode and the reservoir. For a small ring diameter the spatial overlap of the mode with the reservoir is large, the mode is quickly populated, but also decays fast. When the spatial overlap reduces with increasing ring diameter, the mode is not populated as fast and also decays slower. Accordingly, the whole process is slowed down, the peak shifts to later times and also strongly broadens in time. So, the ring diameter of the excitation beam strongly influences the decay time of the zero OAM mode.
	
	The other findings concerning the ring diameter dependence of the -1 and +1 OAM modes can be explained as follows:
	The physical size of vortex modes is influenced by the excitation power density. Here the excitation power and ring diameter are changed simultaneously, so that the peak power density of the excitation ring on the sample and thereby also the predominant vortex size are mostly fixed. When the ring trap diameter is smaller than the vortex modes, only the zero OAM mode can form, since the potential of the trap pushes away the polaritons, which potentially could form vortices. To form a vortex mode the diameter of the trap needs to be a bit larger than the actual vortex size, so that the polaritons can flow freely and form vortices.
	With a ring diameter slightly bigger than the size of the vortex modes (as seen from Fig. \ref{fig:VortexResult}), localized vortices, which have a large spatial overlap with the ring shaped reservoir, may form spontaneously. Due to the strong overlap with the reservoir, the modes form quickly, but also decay fast, since the reservoir is depleted quickly as well. 
	When the ring diameter is much bigger than the size of the vortices, the vortex modes still are created, but may move inside the trap, for example due to an asymmetric intensity distribution of the pump ring or the disorder potential of the sample. When a vortex moves away from the central position in the trap, this also changes the OAM with respect to this central position. Since our detection scheme using the OAM sorting measures the OAM with respect to a fixed axis at the centre of the ringlike trap, the centre of mass motion of vortices appears as a small shift of the OAM modes away from the integer values. So, we conclude that the tilt of the spots in the right panel of Fig. \ref{fig:Different_Rings} indeed corresponds to spontaneously created vortices, which both are moving deterministically into the same direction inside the trap. This result also explains the enhanced life time of the -1 and +1 OAM modes. When the vortices move away from the centre of the trap, the spatial overlap with the feeding reservoir is increased in some areas and reduced in some other areas, leading to a spatially inhomogeneous feeding rate of the reservoir into the vortex mode. The slower feeding rates from the local reservoir areas with reduced spatial overlap thereby increase the total life time of the vortex modes.
	
	\section{Conclusion}
	To conclude, OAM spectroscopy has practical advantages compared to interferometric techniques and enables easily conductable studies of optical vortices. We find that different OAM modes can be generated using an annular non-resonant pumping pattern created by short pulses with a ring-shaped intensity profile.
	Vortices of both positive and negative topological charges are created spontaneously. These modes with +1 and -1 OAM are independent of each other, as indicated by an extensive correlation analysis. Furthermore also a condensate with zero OAM forms at a later time. The excitation power and ring diameter shift all modes with and without OAM to earlier times with increasing excitation power and decreasing ring diameter. We provide a theoretical model based on an active and inactive reservoir, which is in good qualitative agreement with our experimental results and highlights the importance of the excitation geometry for the condensate formation. While choosing ring diameters smaller than the vortex modes prevents the formation of vortices, localized vortices can be created by adjusting the trap diameter to be only slightly bigger than the vortex itself. With a trap diameter significantly larger than the $|m|=1$ vortices, the effective measured OAM shifts in time, indicating a movement of the vortices inside the trap.	
	So the trapping potential essentially given by excitation power and ring diameter has a strong influence on the localization and stability of the vortex modes.
	
	\section{Acknowledgments}
	
	We gratefully acknowledge financial support by the Deutsche Forschungsgemeinschaft (DFG) through the international collaborative research center TRR142 (grant No.  231447078, project A04) and Heisenberg program (grant No. 270619725) and by the Paderborn Center for Parallel Computing. X.M. further acknowledges support from the NSFC (No. 11804064). The Würzburg group acknowledges support by the state of Bavaria.

\end{document}